# Observation of the topological edge state in X-ray band

Qiushi Huang[*], Zhiwei Guo[*], Jiangtao Feng, Changyang Yu, Haitao Jiang[†], Zhong Zhang, Zhanshan Wang[†], and Hong Chen

*Key Laboratory of Advanced Micro-structure Materials, MOE, School of Physics Science and Engineering, Tongji*

*University, Shanghai 200092, China*

## Abstract

The possibility of obtaining robust edge state of light by mimicking the topological properties of solid state system, have brought a profound impact on optical sciences. With the advent of high-brilliance, accelerator-driven light sources such as storage rings or X-ray lasers, it has become attractive to extend the concept of optical topological manipulation to the X-ray regime. In this paper, we theoretically proposed and experimentally demonstrated the topological edge state at the interface of two photonic crystals having different band-gap topological characteristics for X-ray. Remarkably, this topologically protected edge state is immune to the weak disorder in form of the thickness disorder and strong disorder in form of the positional disorder of layers in the structure, as long as the zero-average-effective-mass condition is satisfied. Our investigation therefore brings the topological characteristics to the X-ray regime, provides new theoretical tools to study X-ray optics and may pave way to exploit some important potential applications, such as the high efficiency band filter in X-ray band.

[*] These authors contributed equally to this work

[†] jiang-haitao@tongji.edu.cn

[†] wangzs@tongji.edu.cn

# I. INTRODUCTION

The explorations of topological properties and robust chiral transport have been the subject of intensive research in many braches of physics, from electronics [1, 2] to photonics [3-25], acoustic and phononics [26-29] as well as mechanics [30, 31]. Photonic topological insulators that can manipulate light in a manner similar to the topological insulators controlling the electrons are rapidly developing in the past few years. The early studies are focused on photonic systems with relatively large physical size in the low-frequency micro-wave band, including gyromagnetic materials [5, 6], bi-anisotropic metamaterials [7, 22], and coupled resonator waveguides [9, 10, 19]. Recently, based on dielectric photonic crystals (PCs), photonic topological insulators have been realized in visible band and used for light splitting, unidirectional propagation, etc. [8, 25]. The most remarkable property of photonic topological insulators is that there would be some kinds of topological protection for the light propagation. As a result, photonic topological insulators facilitate the design of novel photonic devices that are robust against the disorder and fluctuations [7, 17, 24]. On the other hand, the short wavelength optical science, particularly in the X-ray region, is undergoing a tremendous development owing to its unique advantages of high resolution in microscopy, spectroscopy and ultrafast dynamics studies [32-35]. With the advent of a new generation of high-brilliance, high coherent X-ray sources, such as the lab-based plasma source, high harmonic generation, and the accelerator-based large facilities, the advantages of X-ray science and technology are further promoted. To materialize the powerful X-ray tool, novel optics with flexible functionalities are demanded to reflect, focus and shape the X-ray beam in an exquisite way [36, 37]. Beside the application need, modern thin film deposition and nanofabrication technology is also advancing rapidly and one-dimensional (1-D) and two-dimensional (2-D) nanoscale structures with large area have been frequently realized. For example, ultra short period multilayer mirrors with periodic bilayer thickness of 1.6~2.5 nm are developed for soft X-ray water window microscopes [38] or hard X-ray monochromators for synchrotron beam lines [39], in order to select the desired waveband. In addition, aperiodic multilayer mirrors are frequently produced to achieve broadband angular or spectral response [40]. Two dimensional 15 nm particle arrays have also been fabricated in centimeters area using photolithography [41]. For X-ray devices working in the very short wavelength, the fabricating difficulty is much higher than those working in the visible wavelength. The fabricating accuracy for X-ray devices should be very high and even small perturbations would strongly degrade the performance of the devices. Therefore, given to the wide applications and significant progress of technology for X rays, a question naturally arises: can the topological properties be applied in the X-ray region and enable new phenomena and devices that are robust against perturbations? In this paper, we use a 1-D symmetric PC system to demonstrate the topological band-gaps and the edge states in X-ray band for the first time. Very interestingly, these topologically protected edge states are robust against a variety of perturbations, such as thickness disorder (weak disorder) and positional disorder (strong disorder), which can facilitate the design of novel X-ray filters/mirrors immune to some kinds of disorders.

In general, the edge state can be formed at the boundary separating two PCs having different band-gap topological invariant, such as geometric phase in 1-D systems [14, 27]. As a geometric phase initially acquired from the electron Bloch state, the Zak phase is considered to the a $Z_2$ topological invariant ($\pi$ or 0) for classifying bands of 1-D electronic systems with inversion symmetry [42]. By analogy, for the 1-D photonic system with inherent mirror symmetry, the corresponding Zak phase is a quantized value of 0 or $\pi$ [27, 42]. In particular, the Zak phase of the bands can be directly determined by calculating the reflection phase of the band-gap [43]. Moreover, by mapping Maxwell's equations to the Dirac equation, the band-gaps of 1-D PCs with different topological orders can be mapped to the band-gaps with effective negative permittivity or negative permeability [4, 13]. Here, we will verify the topological nontrivial insulators in X-ray band from two kinds of reflection phases and the effective parameters. Considering the interest of fundamental science, we firstly designed and validated the topological edge state in X-ray band. Actually, almost all the natural material can hardly reflect the X wave and it cannot be used as a reflector for X wave. As a good solution, the X-ray reflector is constructed by utilizing band-gap of multilayer structures to reflect the X-rays. To achieve designed property, periodic structures with the unit-cell size comparable with or even smaller than the wavelength are needed. As a result, the performance of the multilayer structures is often degraded by the fluctuation of the layers and the difficulty of designing X-ray band-gap is how to deposit the nano-meter scale thick multi-layer with layer thicknesses typically oscillating in the range of 1.0-5.0 nm [44, 45]. Here we introduce the concept of topology into X-ray band and realize the X-ray topological edge state. At this topological edge state, we experimentally demonstrate that the transmission property does not undergo noticeable degradation with moderate degree of disorders, which may relieve the strict tolerance in the fabrication of X-ray filters.

The paper is organized as follows. In Sec. II, by calculating the reflection phases and the effective parameters, we reveal that two photonic insulators with different topological properties for X-ray band can be realized in 1-D symmetry PCs. Moreover, we demonstrated that the topological edge state forms at the boundary separating two PCs having different band-gap topological characteristics for X-ray. In Sec. III, we demonstrated experimentally that the topologically protected edge state is robust against a variety of perturbations, such as thickness disorder and positional disorder. Finally, we conclude in Sec. V.

## II. THE DESIGN OF THE TOPOLOGICAL EDGE STATE FOR X-RAYS

We begin with the design of the two 1-D symmetry PCs for X-ray band with different topological properties. This is because the topological property of the symmetric structure is easier to characterize than the asymmetric structure [4, 14, 27]. Two different PCs are marked by $PC_A$ and $PC_B$, respectively, as is shown in Fig. 1. The unit cells of the PCs are shown in the inset of Fig. 1. It is generally known that the refractive indices of materials are close to 1 in X-ray regime. This feature is very unfavorable to design band-gap for PC and to obtain enough reflectance it must be grazing incident to the PC. Considering the factors of high refractive index difference, low loss, thin film thickness and the experimental processing technology, the layer $A_1$ and $A_2$ in $PC_A$ are carbon (C) and tungsten (W) with $d_c = 1.5nm$, $d_w = 1.5nm$. These materials have relatively large refractive indices contrast in the frequency range of interest. When the frequency is fixed to $1.94 \times 10^{18}\ Hz$, the refractive indices are [46]:

$$\begin{array}{l} n_C \approx 0.99999 + 1.11473 \times 10^{-8} \\ n_W \approx 0.99995 + 3.94227 \times 10^{-6} \end{array}. \quad (1)$$

The structure $PC_A$ is denoted by $(CWC)_{10}$ with the thickness of the unit cell and the period of the unit cell are $d_A = 4.5nm$ and $N_A = 10$, respectively. In general, the band-gap realized by PC can be tuned to be closed and reopened by changing the materials or the thickness of the different layers. Moreover, there is a topological transition in this process [14, 27]. So the $PC_B$ with distinct topological properties to $PC_A$ can be easily realized by changing the layers of $PC_A$. In order to simplify the experimental operation, the layer $B_1$ and $B_2$ in $PC_B$ are tungsten (W) and carbon (C) with $d_W = 1.5nm$, $d_C = 1.5nm$, respectively. The structure $PC_B$ is denoted by $(WCW)_{10}$ with the thickness of the unit cell and the period of the unit cell are $d_B = 4.5nm$ and $N_B = 10$, respectively.

Take the first structure (be marked as $PC_A$) for example, which consists of alternative layers of $A_1(n_{A1})$, $A_2(n_{A2})$ and $A_1(n_{A1})$ with the thickness of $d_{A1}$, $d_{A2}$ and $d_{A1}$, respectively. $n_{A1}$ and $n_{A2}$ are the refractive indices of $A_1$ and $A_2$ layer, respectively. The thickness of the unit cell is $d_A = d_{A1} + d_{A2} + d_{A1}$. The period number of unit cell is $N_A$. To demonstrate the topological differences of the 1-D PCs in X-ray band, we have calculated the reflection phase and effective parameters for the designed PCs. Two different PCs are working at the frequency $1.94 \times 10^{18}\ Hz$ (Cu-Kα line E = 8.04 kev) in the grazing incidence range of $\alpha \in (0.8^o,\ 1.3^o)$.

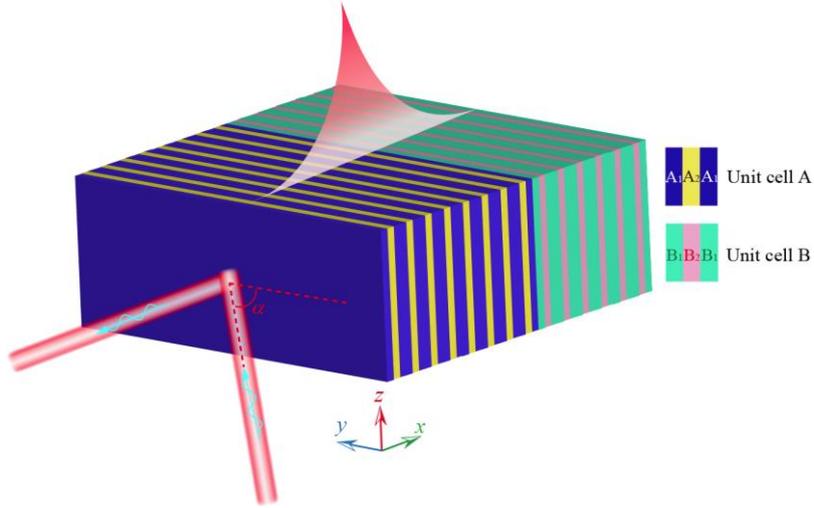

Fig. 1. The scheme of the topological edge mode formed at the interface separating two PCs having different bandgap topological characteristics when the X-ray impinges on the sample at oblique incidence. $PC_A$ is on the left-hand of the interface and $PC_B$ is on the right-hand of the interface. The corresponding unit cells of two different PCs are shown in the inset.

Next, we analyze the topological properties of $PC_A$ and $PC_B$. Particularly, the physical mechanism of the formation of the topological edge state and the corresponding electric field distribution are verified. Firstly, the effective parameters of $PC_A$ and $PC_B$ can be get by the optimal effective medium theory [4, 47], which are shown in Fig. 2(a) and Fig. 2(b), respectively.

By comparing the Dirac equation of electronic system and the Maxwell equation of photonic system, the Maxwell equations can be written as the form of Dirac equation [13]

$$[-i\sigma_x \partial_x + m(x)\sigma_z + V(x)]\begin{pmatrix}\sqrt{\varepsilon_0}E_z \\ \sqrt{\mu_0}H_y\end{pmatrix} = E\begin{pmatrix}\sqrt{\varepsilon_0}E_z \\ \sqrt{\mu_0}H_y\end{pmatrix}, \qquad (2)$$

where $m(x) = \left(\frac{\omega}{2c}\right)[\varepsilon_r(x) - \mu_r(x)]$ denotes the effective mass. $\omega$ is the angle frequency. $V(x)$ and $E$ are the effective potential and energy eigenvalue, respectively [13]. Considering the X-ray band-gap of $PC_A$, which can be effective to the epsilon-negative (ENG) metamaterial with negative permittivity and positive permeability, the effective mass is a negative value. However, for the X-ray band-gap of the structure $PC_B$, which can be effective to the mu-negative (MNG) metamaterial with negative permeability and positive permittivity, the effective mass is a positive value. According to the Dirac equation, the edge state exists in the hetero-structure composed of positive and negative masses when $\bar{m} = \int_{-L_1}^{0} m_1(x)dx + \int_{0}^{L_2} m_2(x)dx = 0$, where $m_1(x)$ and $m_2(x)$ are effective masses of the length $L_1$ and $L_2$, respectively [48]. Moreover, the sign of effective mass corresponds to different topological order [48]. For the Cu-Kα line (λ=0.154 nm) $1.94 \times 10^{18}$ $Hz$, there is an edge mode exist in the angle spectrum when the incident grazing angle is α = $1.07°$. In this case the effective masses of $PC_A$ and $PC_B$ are $m_A \approx -2.03 \times 10^8$ and $m_B \approx +2.03 \times 10^8$, respectively. It has shown that edge state in 1-D system is robust against effective mass distribution so long as $\bar{m} = m_A + m_B = 0$ [4, 13]. According to the relationship between effective parameters and the topological invariant [4, 13], we can conclude the band-gaps of $PC_A$ and $PC_B$ can be seen as two X-ray insulators with different topological properties and the edge state will appear at the interface of two topological distinct structures.

In order to the further verify the topological difference between $PC_A$ and $PC_B$, we have also calculated the reflection phase, which is another quantity that is considered to be able to determine the topological property of the band-gap [27]. The reflectance phases of $PC_A$ and $PC_B$ belong to $(-\pi, 0)$ and $(0, \pi)$, respectively, which are shown in Fig. 2(c). So, we can conclude that these two PCs are topologically distinct. Take into account the losses of C and W layers in $PC_A$ and $PC_B$, the incident X-ray will be partially reflected on the incident plane rather than the perfect tunnelling. But in any way, the topological edge state can be localized at the interface of two different topologically distinct structures, as is shown in Fig. 2(d). Moreover, the topological edge states are protected by the topological phase transition across the interface. Thus they are robust against the defects, the loss, and the disorder, which do not change the topological phase of the structure [17, 24].

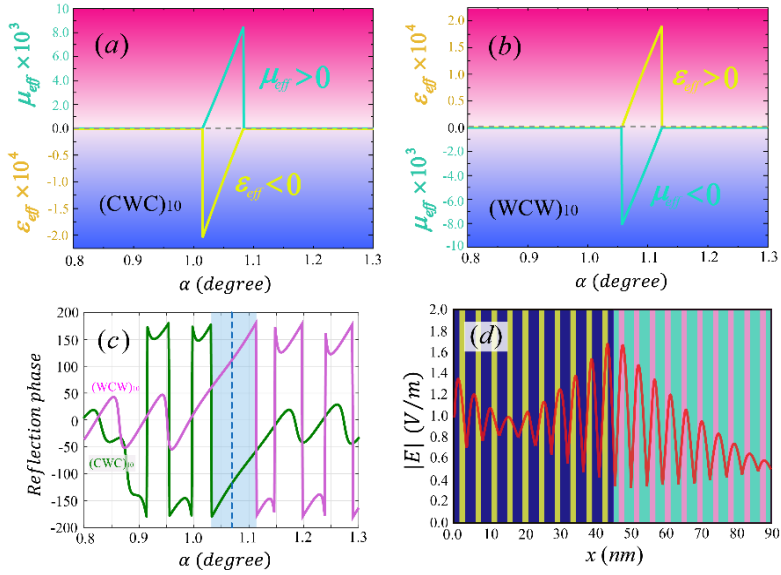

Fig.2 The separate structructures $PC_A$ and $PC_B$ are corresponding to the ENG (a) and MNG metamaterials (b), respectively. (c) The reflection phases of $PC_A$ (marked by the green line) and $PC_B$ (marked by the purple line), respectively. (d) The electric field distribution in the composite structure $PC_A - PC_B$. The edge mode is formed at the boundary between two topologically distinct structures $PC_A$ and $PC_B$. The reflection at the entrance face of the structure is mainly caused by the losses of C and W layers.

The reflectance of grazing incidence X-ray is performed using Bede Refs software and IMD software [48]. When the X-ray is incident to the $PC_A$, there is a band-gap (painted in green ) in the angular spectrum, as is shown in Fig. 3(a). Similarly, there is also a band-gap in the angular spectrum (painted in pink ) when the X-ray is incident to $PC_B$, see Fig. 3(b). It can be found that when the X-ray is incident to the hetero-structure composed of $PC_A$ and $PC_B$, the reflection dip will appear in the middle of the band-gap, as is shown in Fig. 3(c). In other words, the separate $PC_A$ and $PC_B$ are the insulators of X-ray. However, when the X-ray impinges on the structure combined of $PC_A$ and $PC_B$, the structure become nearly transparent because the topological edge state will form at the boundary separating two PCs, which corresponds to the reflection dip $\alpha = 1.07^o$ in Fig. 3(c). It should be noted that this topological edge mode is not limited at the fixed frequency $1.94 \times 10^{18}\ Hz$ and it can be tuned in a certain frequency scope within the bandgap

The reflectance of the edge state in Fig. 3(c) is not zero, which is caused by the losses of C and W layers in $PC_A$ and $PC_B$. Considering the three structures mentioned above, we have calculated the reflectance for different frequency and the grazing incident angle, which is shown in Fig. 3(d)-(f). The reflection angle spectra of Fig. 3(a)-(c) correspond to the special case ($1.94 \times 10^{18}\ Hz$) of Fig. 3(d)-(f), which are marked by the white dashed lines.

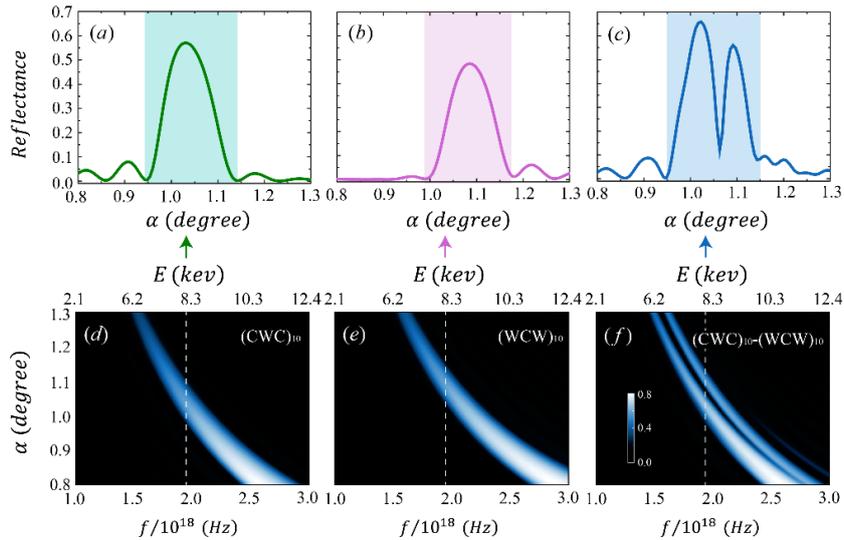

Fig.3 (a), (b), (c) The reflection angle spectra of the separate structructures $PC_A$, $PC_B$ and the composite structure $PC_A - PC_B$, respectively.(d), (e), (f) The reflectance for the different frequency and the grazing incident angle. The white dashed lines marked in the picture indicate the position of $1.94 \times 10^{18}\ Hz$.

# III. EXPERIMENTAL VERIFICATION OF THE TOPOLOGICALLY PROTECTED EDGE STATE

In this section, we experimentally demonstrate our theoretical design of topological edge state in X-ray by fabricating and measuring the corresponding 1-D PC structures. The 1-D PCs with and without disorders are fabricated by direct current magnetron sputtering technique [49, 50], owing to the extremely small layer thickness. High purity tungsten (99.95%) and carbon (99.999%) are used as sputtering targets. The base pressure before deposition was ~$2\times10^{-4}$ Pa and the sputtering gas is argon with the pressure of 0.13 Pa. The deposition rate of W and C are controlled to be very small, 0.07 nm/s and 0.03 nm/s, in order to grow the structure accurately. Super-polished silicon wafer is used as the substrates with the root-mean-squared surface roughness of 0.2 nm. After deposition, the standard $PC_A - PC_B$ sample is measured with transmission electron microscope (TEM) to check the structure of the composite layered system and the layer growth quality. The X-ray reflectance of all samples are measured by a commercial X-ray diffractometer (Bede D1) at the Cu-Kα emission line ($1.94 \times 10^{18}\ Hz$). The measurements are performed by the standard θ-2θ scan, while θ is the grazing incident angle (angular position of the sample) and 2θ is the angular position of detector. Here, θ represents the same angle as α defined above. The angular divergence of the incident beam is 0.007° and the scanning angular step is Δθ=0.005°.

The TEM image of the fabricated $PC_A - PC_B$ sample is shown in Fig. 4(a). The bright layers are C and the dark layers are W. The transition area between the two PCs is measured and magnified displayed in Fig. 4(b). The nano-layers are grown with smooth and sharp interfaces which indicate a good quality of the multilayers. For the $PC_A$ structure ($(CWC)_{10}$), the carbon layers from neighboring unit cells grow continuously forming a thicker carbon layer as seen in the top part of the image. For the PC$_B$ structure ((WCW)$_{10}$), the tungsten layer looks thicker for the same reason. To further determine the layer thicknesses, line profiles are drawn on the cross-section as shown in Fig. 4(c). According to the variation of the gray-scale values, the W layers in each unit cell are estimated to be ~1.7 nm and C layers are ~ 1.4 nm. The thicknesses are very close to the designed structure. A small inter-diffusion region of 0.4-0.5 nm width is found at the interfaces between W and C. Figure 4(d) shows the measured reflectance profiles of the $PC_A$, $PC_B$, and the composite $PC_A - PC_B$ samples, respectively. We can clearly see that separate $PC_A$ and $PC_B$ exhibit a typical Bragg peak of the mirrors. However, once the topologically distinct $PC_A$ and $PC_B$ are combined together, there is a reflection dip in the X-ray band-gap, which is predicted by the topological edge state. The experimental results in Fig. 4(d) well coincide with the theoretical simulation in Fig. 2(c) except that the angular position of the edge state shifts a little to the small angle side, owing to the slightly larger thickness of the deposited layers. The agreement between the test and the theory shows that the topological edge state model is applicable to the X-ray band.

According to the theory described in the previous part II, the topological interface state will exist as long as the zero-average-effective-mass ($\bar{m} = 0$) condition is satisfied. In the next, keeping the structure of $PC_A$ unchanged and adding some kinds of disorders into $PC_B$, we have revealed that the X-ray edge state in the hetero-structure is robust against a variety of disorders. At first, for the thickness disorder of the layers, we have demonstrated that the topological edge is maintained for the weak disorder when $\bar{m} = 0$ is satisfied. The first structure of $PC_B^{'}$ with weak disordered is marked as $(B_1B_2B_1)_1(B_1B_3B_1)_1(B_1B_4B_1)_2(B_1B_3B_1)_1(B_1B_2B_1)_2(B_1B_3B_1)_1(B_1B_2B_1)_1(B_1B_4B_1)_1$,

where the layer $B_3$ and $B_4$ are the carbon layer with thickness 2 nm and 1 nm, respectively. Subscripts represent the number of the unit cell. In this case, the edge mode in the hetero-structure $PC_A - PC_B'$ is maintained as the case without disorders because the effective-mass of $PC_B'$ is still

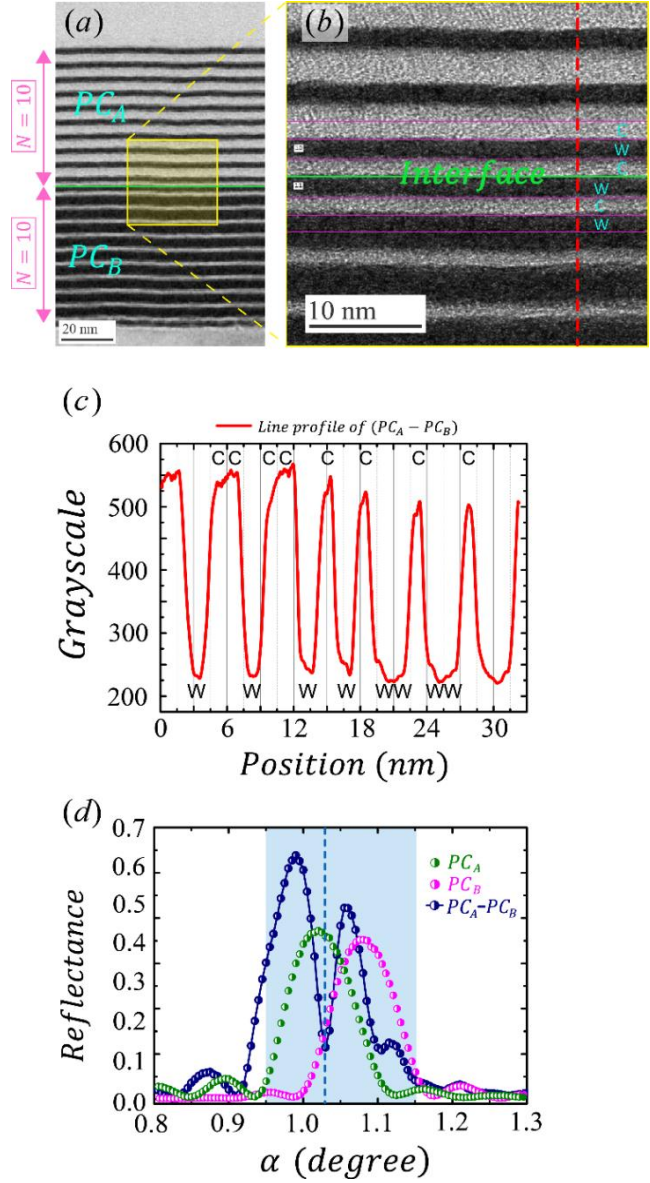

Fig. 4(a). TEM image of the sample. Enlarged picture is shown in the inset picture, here the interface of $PC_A$ and $PC_B$ is marked by the green line. (b). Zoomed-in photo of the dashed region in (a). (c). Line profile of $PC_A - PC_B$ along the red dashed line marked in (b). (d). The reflection angle spectrums of the separate structructures $PC_A$, $PC_B$ and the composite structure $PC_A - PC_B$ measured by experiment. The combine of two kinds PCs is marked by the shaded area.

$m_B \approx +2.03 \times 10^8$ and $\bar{m} = 0$, as is shown in Fig. 5(a). The weak disordered samples are fabricated while part of the carbon layers are intentionally deposited to 2 nm and 1 nm according to the design. The experimental measured reflectance is shown in Fig. 5(b) and the topological edge mode of the $PC_A - PC_B'$ with $\bar{m} = 0$ remains almost the same as the standard structure. The experiment results are in good agreement with the theoretical results. Furthermore, the reflection peaks near the topological edge state also are maintained for the weak disordered case. This feature is very useful in the application of X-ray filter. Secondly, in the case of $\bar{m} \neq 0$, the edge mode is significantly distorted as predicted by the theory. Considering the second weak disordered case that the $PC_B''$ is marked as $(B_1B_4B_1)_2(B_1B_2B_1)_1(B_1B_4B_1)_2(B_1B_3B_1)_1(B_1B_2B_1)_2(B_1B_3B_1)_1(B_1B_2B_1)_1$, the X-ray edge state in the structure $PC_A - PC_B$ will be influenced obviously, because the effective-mass of $PC_B''$ is not equal to $2.03 \times 10^8$ and $\bar{m} \neq 0$, as are shown in Fig. 5(c) and 5(d). So, we demonstrate experimentally that the topological edge state is protected by the $\bar{m} = 0$ condition.

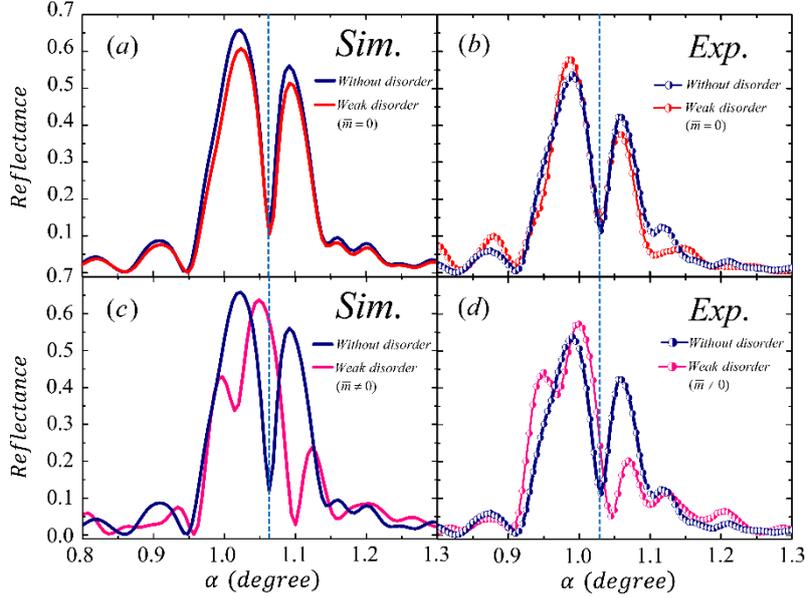

Fig. 5(a), (b) For the weak disorder (thickness disorder) with $\bar{m} = 0$, the X-ray edge mode is hardly affected. (c), (d) For the weak disorder (thickness disorder) with $\bar{m} \neq 0$, the X-ray edge mode is greatly influenced.

At last, different from the type of weak disorder added into the structure, in this part, by keeping the thickness of the layers invariant, the robustness of the topological edge state is further examined by adding strong disorder (positional disorder) in the hetero-structure $PC_A - PC_B'''$ as

$(A_1A_2A_1)_2(B_1B_2B_1)_2(A_1A_2A_1)_3(B_1B_2B_1)_2(A_1A_2A_1)_1(B_1B_2B_1)_4(A_1A_2A_1)_1(B_1B_2B_1)_2(A_1A_2A_1)_3$ .

Although this situation corresponds to a strong disorder case, the topological edge is robust against the perturbation because the condition of $\bar{m} = 0$ is satisfied, as is shown in Fig. 6. Different from the weak disordered structure introduced in Figs. 5(a) and 5(b), although the topological edge state is still maintained in the case of strong disordered structure in Fig. 6, the reflection peaks near the edge state are obviously suppressed. The whole reflectance curve is very close to the designed result of the corresponding structure which implies an accurate fabrication process. The edge mode remains almost at the same position as the standard structure and the small shift is caused by the little thickness deviation of the composing unit cells between the two samples. It is demonstrated again that, even for strong disorders, the $\bar{m} = 0$ condition can still protect the edge mode against randomness, which is very useful in the applications of X-rays.

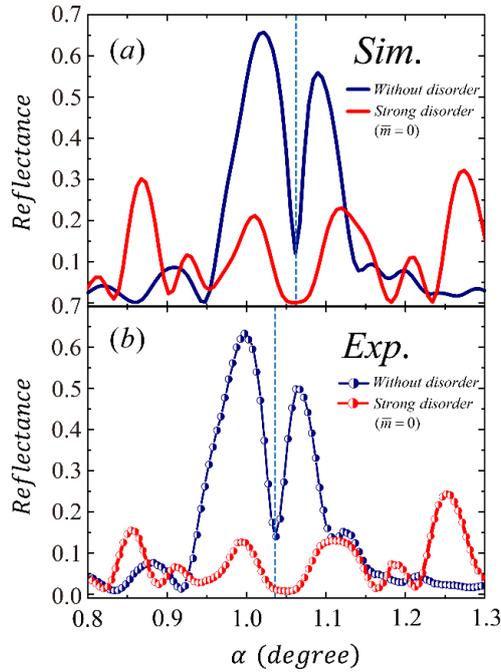

Fig. 6 The robustness of X-ray edge mode against the strong disorder with $\bar{m} = 0$ has been demonstrated by simulation (a) and experiment (b), respectively.

In discussion, we firstly extend the concept of topology to the X-ray band. The distinct topological properties of two different PCs have been revealed from the effective parameters and the reflection phases. The predicted topologically protected edge states at the interface of two PCs with different topological properties also have been demonstrated. When a variety of disorders are introduced into the system, the topological edge state can always be maintained as long as the $\bar{m} = 0$ condition is satisfied. All the results have been verified by the experiments that are in good agreement with the theory. Our results will provide new theoretical tools to study X-ray optics and may provide a possibility for the development of new X-ray devices.

## IV. CONCLUTION

In summary, topological edge state in X-ray band has been theoretically proposed and experimentally demonstrated for the first time based on two 1-D PCs with different topological properties. This edge state is robust against a variety of perturbations, such as thickness disorder (weak disorder) and positional disorder (strong disorder), as long as the $\bar{m}=0$ condition is satisfied. Our system provides a photonic platform for exploring X-ray propagations and developing possible topological devices. Our results may promote the development of X-ray optics by introducing the new topological freedom. Although our results are based on 1-D system, the concept of topological manipulation might be consulted for higher dimensional system and related X-ray devices. Therefore, our investigation not only extends the range of applicability of the photonic topological theory, but also provides insightful guidance to exploring the exciting applications associated with topological transport to higher spectral range.

**APPENDIX**

For comparison, different geometrical structures with and without disorders are given in Table 1. The materials and thicknesses of different layers are listed in Table 2. At last, Table 3 shows the relationship between the $\bar{m}=0$ condition and the topologically protected edge state immune to the disorders.

**Table 1.  Different structures**

| Structure | Layer parameters |
|---|---|
| 1 | $(A_1A_2A_1)_{10}(B_1B_2B_1)_{10}$ |
| 2 | $(A_1A_2A_1)_{10}(B_1B_2B_1)_1(B_1B_3B_1)_1(B_1B_4B_1)_2(B_1B_3B_1)_1(B_1B_2B_1)_2(B_1B_3B_1)_1(B_1B_2B_1)_1(B_1B_4B_1)_1$ |
| 3 | $(A_1A_2A_1)_{10}(B_1B_4B_1)_2(B_1B_2B_1)_1(B_1B_4B_1)_2(B_1B_3B_1)_1(B_1B_2B_1)_2(B_1B_3B_1)_1(B_1B_2B_1)_1$ |
| 4 | $(A_1A_2A_1)_2(B_1B_2B_1)_2(A_1A_2A_1)_3(B_1B_2B_1)_2(A_1A_2A_1)_1(B_1B_2B_1)_4(A_1A_2A_1)_1(B_1B_2B_1)_2(A_1A_2A_1)_3$ |

**Table 2.  Materials and the thicknesses of different layers**

| Layer | $A_1$ | $A_2$ | $B_1$ | $B_2$ | $B_3$ | $B_4$ |
|---|---|---|---|---|---|---|
| Material | C | W | W | C | C | C |
| Thickness (nm) | 1.5 | 1.5 | 1.5 | 1.5 | 2.0 | 1.0 |

**Table 3.  Relationship between the $\bar{m}=0$ condition and the topologically protected edge state**

| Structure | Disorder | Effective mass | Exist edge mode |
|---|---|---|---|
| 1 | Without | $\bar{m}=0$ | Yes |
| 2 | Weak | $\bar{m}=0$ | Yes |
| 3 | Weak | $\bar{m}\neq 0$ | No |
| 4 | Strong | $\bar{m}=0$ | Yes |

ACKNOWLEDGMENT

This work is supported by the National Key R&D Program (No. 2016YFA0301101, 2016YFA0401304, 2017YFA0403302); by the National Nature Science Foundation of China (NSFC) (Grant Nos. 11774261, 11474220, 61621001, 11505129, and U1732268); the National Natural Science Foundation of China Academy of Engi-neering Physics (NSAF) (No. U1430131), Science Foundation of Shanghai (No. 17ZR1443800); We thank Kun Yu, Feng Wu, and Yang Yang for helpful discussions.